\documentclass[a4paper,11pt]{article}
\pdfoutput=1 

\usepackage{jcappub} 

\usepackage[T1]{fontenc} 
\usepackage{newtxtext,amssymb,amsmath,extarrows}
\title{\boldmath Cosmological Collider Signatures of Massive Vectors from Non-Gaussian Gravitational Waves}

\author{Yi Wang}
\author{and Yuhang Zhu}

\affiliation[1]{Department of Physics, The Hong Kong University of Science and Technology,\\Clear Water Bay, Kowloon, Hong Kong, P.R. China}
\affiliation[2]{The HKUST Jockey Club Institute for Advanced Study,\\The Hong Kong University of Science and Technology,\\Clear Water Bay, Kowloon, Hong Kong, P.R. China}

\emailAdd{phyw@ust.hk}
\emailAdd{yzhucc@connect.ust.hk}

\abstract{
	The cosmological collider provides a model-independent probe of particle physics during inflation. We extend the study of cosmological collider physics to much smaller scales through gravitational wave (GW) probes. With a Chern-Simons interaction, a massive vector field can obtain a chemical potential and its particle production can cause significant non-Gaussian GW signals. We calculate the mass and spin dependences of the induced GW 3-point correlation function in the squeezed limit, and estimate its amplitude. Such signals may be detectable in the current and upcoming GW interferometer experiments.
}

\begin{document}
\maketitle
\flushbottom

\section{Introduction}
\label{sec:intro}
Inflation is the leading paradigm to describe the very early stage of our universe, which provides a simple mechanism to solve the flatness and horizon puzzles of the hot big bang cosmology. In addition, inflation predicts primordial fluctuations which become the seeds of the fluctuations in the Cosmic Microwave Background (CMB) and Large Scale Structure (LSS). 

The inhomogeneity of the universe can be studied through the connected n-point correlation functions of fluctuations. The three- or higher-point correlation functions are called non-Gaussianities. Among different possible non-Gaussian signals, the squeezed limit  ($k_1 \approx k_2\gg k_3$) three-point function is an especially informative channel. From the oscillatory signature in the squeezed limit, one can read off signatures of additional massive fields during inflation. More specifically, the massive fields with spin $s$ and mass  $m$ can generate an oscillatory signature $(k_3/k_1)^{i\mu_s}P_s{(\cos\theta)}$, where $\mu_s \equiv \sqrt{(M^2/H^2)-\alpha^2}$ and $\alpha$ depends on the spin of massive fields. In this way, the mass and spin information can be extracted from the cosmological correlation functions, independent of the details of inflation models. This approach is known as the cosmological collider~\cite{cc1,cc2,cc3,cc4,cc5,cc6,cc7,cc8,cc9,cc10,cc11,cc12,cc13,cc14,cc15,cc16,cc17,cc18,cc19,lu,cc21,cc22}. 

The inflationary universe as cosmological collider provides us a possible opportunity to probe new physics beyond the standard model, because the energy scale of inflation is usually believed to be much higher than man-made particle colliders. However, if these fields are too heavy ($m\gg H$), the oscillatory signature is generally suppressed by a Boltzmann factor $e^{-\pi \mu}$, which makes the signal of the very heavy modes extremely hard to detect. 

The Boltzmann factor may be weakened or beaten when more features of inflation is turned on, for example, with features in a potential \cite{Chen:2011zf}, a monodromy \cite{Flauger:2016idt}, or with a higher temperature \cite{Tong:2018tqf}. Recently, it is noted that a chemical potential is a natural mechanism to invalidate the Boltzmann factor, considering that they can be introduced by a dimension-5 operator satisfying the shift symmetry of the inflaton. The impact of chemical potential was studied in~\cite{Barnaby:2010vf,  Barnaby:2011qe,Barnaby:2011vw,Barnaby:2012xt,Sorbo2,sorbo1,Domcke:2016bkh,Cook:2013xea,Linde:2012bt,Meerburg:2012id}. For the Fermion particles, the production is modest due to Pauli exclusion. For the Bosonic case, the production can get exponentially amplified by the factor $e^{\pi\xi}$ where $\xi$ is the chemical potential. For example, in axion inflation model the pseudo-scalar inflaton $\phi$ is generically coupled to U(1) gauge field through the interaction $\mathcal{L}=\frac{\phi}{4f}\tilde{F}_{\mu\nu}F^{\mu\nu}$ which can generate exponentially amplified of gauge particles. In the context of cosmological collider physics, the Fermionic case was studied in~\cite{Chen:2018xck} and the Bosonic case was studied in~\cite{xitong}. For a systematic study, see~\cite{Wang:2019gbi}.

Since the Bosonic particle production is usually not under control, over-produced particles introduce too dramatic effects and is usually tightly constrained if it happened during the first 10 e-folds of observable inflation. However, if the Bosonic particle production process happens at a later stage of inflation closer to reheating, the observational signature may be observable from gravitational waves (GW). For massless vectors, the detectable signature of GW at interferometer scales was proposed in~\cite{sorbo1,Barnaby:2011qe}. In addition, there are some other interesting signatures, such as large tensor non-Gaussianities~\cite{Cook:2013xea} and possible large local non-Gaussianities generated through massive gauge field was studied in~\cite{Meerburg:2012id}. 

In this work, we discuss the massive vector cosmological collider signal from GW generated by particle production during inflation through the interaction between inflaton and massive vector fields. The advantage of using GW is that one can probe very small scale fluctuations. At such small scales, corresponding to late times during inflation, the over production of gauge fields is not tightly constrained and thus no balance (tuning) between mass and chemical potential is needed. In the aspect of observational signatures, the shape of non-Gaussianities of massless gauge field is approximately equilateral as a result of signatures are produced in sub-horizon scales. In contrast, massive vector fields can also have significant signal in squeezed limit which has both non-trivial angular dependence related to the spin and oscillatory behavior related to the mass of the vector.   
This cosmological collider signal is generated through the resonance between massive fields and inflaton which is also amplified by the chemical potential term. Finally, we roughly estimate the amplitude of this signal and show that it is possible to detect through future interferometer experiments, for example, LIGO \cite{Abramovici:1992ah}, Virgo \cite{TheVirgo:2014hva}, KAGRA \cite{Somiya:2011np}, LISA \cite{LISA}, DECIGO \cite{Kawamura:2006up}, Taiji \cite{Hu:2017mde} and Tianqin \cite{Luo:2015ght}.
 
This paper is organized as follows: in Section~\ref{sec1}, we introduce the model and the generation of exponentially enhanced massive vector modes. In Section~\ref{sec2}, we discuss the tensor modes generated by the massive vector fields. In Section~\ref{sec3}, we calculate the bispectrum of gravitons in the squeezed limit and find the cosmological collider signal. We conclude in Section~\ref{conclusion}.

\section{Generation of the massive vector field}\label{sec1}
We consider a pseudoscalar inflaton $\phi$ coupled to a massive vector field through the interaction $\phi F\tilde{F}$. This part is mostly a review of~\cite{lu}. The Lagrangian density is described by
\begin{align}
\mathcal{L}=-\frac{1}{2}(\partial\phi)^2-V(\phi)-\frac{1}{4}F_{\mu\nu}F^{\mu\nu}-\frac{1}{2}m^2A_{\mu}A^{\mu}-\frac{\phi}{4f}\tilde{F}_{\mu\nu}F^{\mu\nu}~,
\end{align}
where $f$ is a constant indicating coupling strength, $F_{\mu\nu}=\partial_{\mu}A_{\nu}-\partial_{\nu}A{_\mu}$ is the field strength and $\tilde{F}^{\mu\nu}\equiv\frac{1}{2}\frac{\epsilon^{\mu\nu\rho\sigma}}{\sqrt{-g}}F_{\rho\sigma}$ with $\epsilon^{0123}=1$.

Although the mass term breaks gauge symmetry, if $A_\mu$ couples to a conserved current\footnote{If $A_\mu$ does not couple to a conserved current, in general we cannot choose $\partial_{\mu}(\sqrt{-g}A^{\mu})=0$. However, the `error' is a longitudinal mode which is not enhanced by a chemical potential, and thus is not important for our purpose.}, one can still differentiate the equation of motion and get $\partial_{\mu}(\sqrt{-g}A^{\mu})=0$, which corresponds to a gauge choice in the $m=0$ case. Then, the equation of motion of the massive vector field can be written as
\begin{align}\label{a_x}
\vec{A}''-\nabla^2 \vec{A}-\frac{\phi'}{f}\vec{\nabla}\times\vec{A}+a^2m^2\vec{A}=0~.
\end{align}
 In momentum space,
\begin{align}
\vec{A}''-\nabla^2 \vec{A}-i \frac{\phi'}{f}\vec{k}\times\vec{A}+a^2m^2\vec{A}=0~.
 \end{align}
 
 Compared with the massless case~\cite{Barnaby:2011vw}, where the vector field has only two modes, now there are three degrees of freedom including two transverse modes and one longitudinal mode. After choosing circular polarization basis, their equations of motion are
 \begin{align}\label{eomA}
 &A^{''}_{\pm}+(k^2+\frac{m^2}{{H^2\tau^2}} \pm \frac{2k\xi}{\tau})A_{\pm}=0 ~,\\    
 &A^{''}_L+(k^2+a^2m^2)A_{L}=0~,
 \end{align}
   where  $\xi=\frac{\dot{\phi}}{2fH}$, and $\dot{\phi}$ represent the rolling speed of the background, which is assumed to be a constant. By noticing that only the transverse polarization part of vector field possibly experiences enhancement or suppression (see~\cite{lu} for detailed discussion) from the chemical potential. Thus, we will ignore the contribution from the longitudinal part in our later calculation. The vector field can be decomposed as
    \begin{align}\label{a_k}
   A_{i}(k,\tau)=\sum_{\lambda=\pm}\epsilon_i^{(\lambda)}A_{\lambda}(\tau,k)a_{\lambda}(k)+h.c.~,
   \end{align}
   and the annihilation/creation operators obey
   \begin{align}
   [a_{\lambda}(\mathbf{k}),a^{\dagger}_{\lambda'}(\mathbf{k'})]=(2\pi)^3\delta_{\lambda\lambda'}\delta^{(3)}(\mathbf{k}-\mathbf{k}')~,
   \end{align}
where $\vec{\epsilon}(k)$ are the polarization vectors, which satisfy the relation $\mathbf{k}\cdot\vec{\epsilon}_{\pm}(\mathbf{k})=0,i\mathbf{k}\times\vec{\epsilon}_{\pm}(\mathbf{k})=\pm k\vec{\epsilon}_{\pm}(k),\vec{\epsilon}_{\pm}(\mathbf{k})\cdot\vec{\epsilon}_{\pm}(\mathbf{k})=0, \vec{\epsilon}_{\pm}(\mathbf{k})\cdot\vec{\epsilon}_{\mp}(\mathbf{k})=1$, and $\vec{\epsilon}_{\pm}(\mathbf{k})=\vec{\epsilon}_{ \pm}(\mathbf{-k})^*$. From the equation of motion of the transverse modes, only one of the two modes would experience a tachyonic instability. Without loss of generality, we can assume that $\dot{\phi}>0$ so that the positive helicity state of the vector field gets copiously produced while the other helicity one is suppressed. In the later discussion, we only consider the contribution from "+" helicity part. The solution is well described by Whittaker functions and after choosing the Bunch-Davies initial condition, the solution of the equation of motion is
 \begin{align}
 A_{+}(\tau,k)=\frac{1}{\sqrt{2k}}e^{\pi \xi/2}W(-i\xi,i\mu,2ik\tau)~,
 \end{align}
 where $\mu^2=\frac{m^2}{H^2}-\frac{1}{4}$. This enhancement effect caused by the chemical potential can be easily seen from the IR ($|k\tau|\rightarrow0$) limit of the mode function \cite{xitong}

\begin{align}\label{vectoreom}
A_{+}(\tau,k)=\alpha \frac{C}{\sqrt{2 \mu}}(-\tau)^{\frac{1}{2}+i\mu}+\beta \frac{C^{*}}{\sqrt{2\mu}}(-\tau)^{\frac{1}{2}-i \mu}~,
\end{align}
where $C=e^{i(\mu \ln(2 k)-\pi/4)}$ and the Bogolyubov coefficients $\alpha,\beta$ are
\begin{align}
\alpha=e^{\frac{1}{2}\pi(\mu+\xi)}\frac{\sqrt{2\mu} \Gamma(-2i\mu)}{\Gamma(\frac{1}{2}-i\mu+i\xi)},\qquad\beta=-ie^{\frac{1}{2}\pi(\xi-\mu)}\frac{\sqrt{2\mu} \Gamma{(2i\mu)}}{\Gamma(\frac{1}{2}+i\mu+i\xi)}~.
\end{align}
For a large $\xi$, the particle number density $\langle n\rangle=|\beta|^2\propto e^{2\pi\xi}$, so that the vector field is exponentially amplified by the chemical potential.

\section{Generation of tensor modes from the massive vector}\label{sec2}
 To obtain the equation of motion of GW sourced by massive vector filed, we write the perturbed metric around the FRW background as
 \begin{align}
 ds^2=a^2(\tau)[d\tau^2-(\delta_{ij}+h_{ij})dx^idx^j]~,
 \end{align}
 where $h_{ij}$ is a transverse ($\partial_ih_{ij}=0$) traceless ($h_{ii}=0$) perturbation of the spatial metric. We have set the scalar as well as vector perturbations to zero. The equation of motion of the perturbation is~\cite{Sorbo2,sorbo1}
\begin{align}\label{eom}
h_{ij}''-\nabla^2h_{ij}+2\mathcal{H}h{'}_{ij}=16\pi GS_{ij}^{TT}~,
\end{align}
where the $S_{ij}^{TT}$ is the transverse-traceless part of the stress tensor 
\begin{align}
S_{ij}^{TT}=(P_{il}P_{jm}-\frac{1}{2}P_{ij}P_{lm})T_{lm}~,
\end{align}
and $P_{ij}=\delta_{ij}-\frac{k_ik_j}{k^2}$ is the transverse and traceless projection operator.
Transform $h_{ij}$ into momentum space
\begin{align}
h_{ij}(x,\tau)=\sum_{\lambda=\pm}\int\frac{d^3k}{(2\pi)^3}\Pi_{ij,\lambda}(k)h^\lambda_k(\tau)e^{-ikx}+h.c.
\end{align}
and
\begin{align}
\Pi_{ij,\pm}(k)=\epsilon_i^{(\pm)}(k)\epsilon_j^{(\pm)}(k)~.
\end{align}
Using the Green's function method, we can write the solution of \eqref{eom} as
\begin{align}
h^{\lambda}_k(\tau)=\frac{2}{Mp^2}\int d\tau'G_k(\tau,\tau')\Pi^{*}_{ij}S_{ij}^{TT}
\end{align}
and $\Pi^*_{ij}(P_{il}P_{jm}-\frac{1}{2}P_{ij}P_{lm})=\Pi^*_{lm}$, so that
\begin{align}\label{hij}
h^{\lambda}_k(\tau)=\frac{2}{Mp^2}\int d\tau'G_k(\tau,\tau')\Pi^{*}_{ij}T_{ij}~,
\end{align}
where the retarded propagator reads~\cite{sorbo1}
 \begin{align}\label{green}
G_k(\tau,\tau')=\frac{1}{k^3\tau'^2}\left[(1+k^2\tau\tau') \sin (k(\tau-\tau'))+k(\tau'-\tau)\cos (k(\tau-\tau'))\right]\Theta(\tau-\tau')~.
\end{align}
The energy-momentum tensor of the produced vector field is
 \begin{align}\nonumber
 T_{ij}&=F_{ia}F^{ia}-\frac{1}{4}g_{ij}F^2+m^2A_iA_j-\frac{1}{2}g_{ij}m^2A_\rho A^\rho\\&=\frac{1}{a^2}\left[-E_iE_j-B_iB_j-\frac{\delta_{ij}}{2}(\mathbf{E}^2+\mathbf{B}^2)\right]+m^2A_iA_j-\frac{1}{2}g_{ij}m^2A_\rho A^\rho~.
 \end{align}
 Since $\delta_{ij}\Pi_{ij}^*=0$,  we can drop the part proportional to the Kronecker $\delta$. Similar to the massless case~\cite{Barnaby:2011vw,Barnaby:2012xt}, the energy density coming from the electric field part dominates over other terms during the production period. To simplify the calculation, in the later discussion only the $E_iE_j$ term is included. Then
 \begin{align}
 T_{ij}(x)=\frac{1}{a^2}[-\partial_{\tau}A_i\partial_{\tau}A_j]~,
 \end{align}
 so that
 \begin{align}\label{tij}
 \Pi^*_{ij,\lambda}(k)T_{ij}(k)&=-\frac{1}{a^2}\int \frac{d^3p}{(2\pi)^3}\Pi^*_{ij,\lambda}(k)\partial_{\tau}A_i(p)\partial_{\tau}A_j(k-p)~.
 \end{align}

\section{The bispectrum of GW}\label{sec3}
In this subsection we calculate the three-point function of tensor fluctuation $h_{\mathbf{k}}$. Combine ~\eqref{hij},~\eqref{green} and~\eqref{tij}, we arrive at 

\begin{align}
\nonumber\langle h_{\mathbf{k_1}}^\lambda h_{\mathbf{k_2}}^\lambda h_{\mathbf{k_3}}^\lambda \rangle&=-\frac{8}{Mp^6}\int\frac{1}{a^2(\tau_1)} d\tau_1G_{k_1}(\tau,\tau_1)\int\frac{1}{a^2(\tau_2)} d\tau_2G_{k_2}(\tau,\tau_2)\int\frac{1}{a^2(\tau_3)} d\tau_3G_{k_3}(\tau,\tau_3)\\&
\times\int \frac{d^3p_1}{(2\pi)^3}\int \frac{d^3p_2}{(2\pi)^3}\int \frac{d^3p_3}{(2\pi)^3}\nonumber\\&\times\langle A'_{i_1}(\mathbf{p_1},\tau_1)A'_{j_1}(\mathbf{k_1-p_1},\tau_1)A'_{i_2}(\mathbf{p_2},\tau_2)A'_{j_2}(\mathbf{k_2-p_2},\tau_2)A'_{i_3}(\mathbf{p_3},\tau_3)A'_{j_3}(\mathbf{k_3-p_3},\tau_3)\rangle~.
\end{align}

Note that we have used a semi-classical formalism for GW. This is not fully consistent with the quantum description of the massive vector. However, considering that at the production period dominated by the chemical potential, the massive vector becomes highly classical, the inconsistency caused by the quantum-classical mismatch is negligible. More explicitly, In the fully quantum in-in formalism, the three-point function is unchanged under exchanging $\mathbf{k_1}$, $\mathbf{k_2}$ and $\mathbf{k_3}$. Here, the order between three momentums will affect the result, because the Whittaker function is a complex function.  However, once the vector field experience the exponentially enhancement period and the situation has transitioned to the classical region. In the IR expansion of the mode function of vector field~\eqref{vectoreom}. The particle number density is $|\beta|^2$ which is proportional to $e^{2\pi\xi}$ and satisfies $|\alpha|^2-|\beta|^2=1$. When chemical potential is sufficiently large, there is almost no difference between $\alpha$ and $\beta$, and the mode function of the vector field becomes a real function up to a constant phase.
Then after Wick contraction, the three-point function becomes
 \begin{align}
&\langle h_{k_1}^\lambda h_{k_2}^\lambda h_{k_3}^\lambda \rangle=\frac{-64}{Mp^6}\int\frac{1}{a^2(\tau_1)} d\tau_1G_{k_1}(\tau,\tau_1)\int\frac{1}{a^2(\tau_2)} d\tau_2G_{k_2}(\tau,\tau_2)\int\frac{1}{a^2(\tau_3)} d\tau_3G_{k_3}(\tau,\tau_3)\nonumber\\&
\times\int {d^3p_1}A'_{+}(p_1,\tau_1)A'^*_{+}(p_1,\tau_2)A'_{+}(k_1-p_1,\tau_1)A'^*_{+}(k_1-p_1,\tau_3)A'_{+}(k_2+p_1,\tau_2)A'^*_{+}(k_2+p_1,\tau_3)\nonumber\\&\times\epsilon_{i_1}^+(\mathbf{p_1})\epsilon_{i_1}^{-\lambda}(\mathbf{k_1})\epsilon_{j_1}^+(\mathbf{k_1-p_1})\epsilon_{j_1}^{-\lambda}(\mathbf{k_1})\epsilon_{i_2}^{-\lambda}(\mathbf{k_2})\epsilon_{i_2}^{+}(\mathbf{-p_1})\nonumber\\&\epsilon_{j_2}^{-\lambda}(\mathbf{k_2})\epsilon_{j_2}^{+}(\mathbf{k_2+p_1})\epsilon_{i_3}^{-\lambda}(\mathbf{k_3})\epsilon^{+}_{i_3}(\mathbf{p_1-k_1})\epsilon_{j_3}^{-\lambda}(\mathbf{k_3})\epsilon^{+}_{j_3}(\mathbf{-k_2-p_1})\delta^3{(\mathbf{k_1+k_2+k_3})}~.
\end{align}
The product term of different polarization vectors is related to 
 non-trivial angular dependence which we will discuss later. For obtaining the oscillation signals in the squeezed limit, we are mostly concerned about the IR behavior of the mode function of massive vector field, 
 \begin{align}
 A'(p,\tau_1)&=\partial_{\tau}A_p\nonumber\\&=-(\frac{1}{2}+i\mu)\frac{\alpha}{\sqrt{2 \mu}} C(p)(-\tau)^{-\frac{1}{2}+i\mu}-(\frac{1}{2}-i\mu)\frac{\beta}{\sqrt{2 \mu}} C^*(p)(-\tau)^{-\frac{1}{2}-i\mu}\nonumber\\&\equiv\alpha_{new}C(p)(-\tau)^{-\frac{1}{2}+i\mu}+\beta_{new}C^*(p)(-\tau)^{-\frac{1}{2}-i\mu}~,
 \end{align}
 where $\alpha_{new}=-\alpha (\frac{1}{2}+i\mu)/(\sqrt{2 \mu})$, $\beta_{new}=-\beta({\frac{1}{2}-i\mu})/\sqrt{2 \mu}$, for simplicity we would  omit the subscript in the following discussion. First of all, the integral about $\tau_1$ is  $A'(p,\tau)A'(\tilde{p},\tau_1)$, where we define $\tilde{\mathbf{p}}=\mathbf{k_1-p}$.

\begin{align}
&A'(p,\tau_1)A'(\tilde{p},\tau_1)=\left(\alpha C(p)(-\tau_1)^{-\frac{1}{2}+i\mu}+\beta C^*(p)(-\tau_1)^{-\frac{1}{2}-i\mu}\right)\nonumber\\&\times\left(\alpha C(\tilde{p})(-\tau_1)^{-\frac{1}{2}+i\mu}+\beta C^*(\tilde{p})(-\tau_1)^{-\frac{1}{2}-i\mu}\right)\nonumber\nonumber\\&=k_1(-k_1\tau_1)^{-1}\left(-i\alpha^2(-k_1\tau_1)^{2i\mu}(\frac{4p \tilde{p}}{k_1^2})^{i\mu}+i\beta^2(-k_1\tau_1)^{-2i\mu}(\frac{4p \tilde{p}}{k_1^2})^{-i\mu}\nonumber+\alpha\beta\left((\frac{p}{\tilde{p}})^{i\mu}+(\frac{p}{\tilde{p}})^{-i\mu}\right)\right)~,
\end{align}
combining this expression with the Green's function part and after changing the variable $-k_1\tau_1$ to $x$, the integral about $\tau_1$ becomes
\begin{align}
I_{\tau_1}&=\frac{H^2}{k_1^3}\int_{0}^{+\infty}\frac{1}{x}\left(\sin x-x\cos x\right)\left[\alpha \beta\left((\frac{p}{\tilde{p}})^{i\mu}+(\frac{p}{\tilde{p}})^{-i\mu}\right)\right]\nonumber\\&+\int_{0}^{+\infty}\frac{1}{x} x^{2i\mu}\left(\sin x-x\cos x\right)\left(-i\alpha^2(\frac{k_1^2}{4p\tilde{p}})^{-i\mu}\right)\nonumber\\&+\int_{0}^{+\infty}\frac{1}{x} x^{-2i\mu}\left(\sin x-x\cos x\right)\left(i\beta^2(\frac{k_1^2}{4p\tilde{p}})^{i\mu}\right)dx~.
\end{align}
For achieving fast convergence in the UV, we will use a more efficient approach of wick rotation~\cite{cc2}. First, translate $\sin$ and $\cos$ into exponential form and then do the wick rotation to achieve the convergence. So that
\begin{align}
\int_{0}^{+\infty}x^{2i\mu} \cos x dx=i\Gamma(1+2i\mu)\sinh (-\pi\mu)~,\\
\int_{0}^{+\infty} \frac{\sin x}{x} x^{2i\mu}dx=i\Gamma(2i\mu)\sinh (\pi\mu)~.
\end{align}
Finally,
\begin{align}
I_{\tau_1}=\frac{H^2}{k_1^3}\left[\frac{\pi}{2}\alpha\beta\left((\frac{p}{\tilde{p}})^{i\mu}+(\frac{p}{\tilde{p}})^{-i\mu}\right)+\alpha^2A\left(\frac{k_1^2}{4p\tilde{p}}\right)^{-i\mu}+\beta^2A^*\left(\frac{k_1^2}{4p\tilde{p}}\right)^{i\mu}\right]~,
\end{align}
where $A=[\Gamma(2i\mu)+\Gamma(2i\mu+1)]\sinh(\pi \mu)$.
Using the same method, the integrals about $\tau_2$ and $\tau_3$ are
\begin{align}
&I_{\tau_2}=\frac{H^2}{k_2^3}\left[\frac{\pi}{2}|\alpha|^2\left(\frac{p}{\tilde{p}}\right)^{-i\mu}+\frac{\pi}{2}|\beta|^2\left(\frac{p}{\tilde{p}}\right)^{i\mu}+\alpha^*\beta A^*\left(\frac{k_2^2}{4p\tilde{p}}\right)^{i\mu}+\alpha\beta^*A\left(\frac{k_2^2}{4p\tilde{p}}\right)^{-i\mu}\right]~,\\
&I_{\tau_3}=\frac{H^2}{k_3^3}\left[\alpha^*\beta^*\pi+(\alpha^*)^2 A^*\left(\frac{k_3^2}{4\tilde{p}^2}\right)^{i\mu} +(\beta^*)^2A\left(\frac{k_3^2}{4\tilde{p}^2}\right)^{-i\mu}\right]~.
\end{align}
The integral about internal momentum $\mathbf{q}$ is still hard to deal with. We can nevertheless estimate the result by estimating the volume of the phase space. We introduce a loop-momentum cutoff $\Lambda \approx k_3$ as in~\cite{Chen:2018xck}, and approximate the measure of the loop integral by
\begin{align}
\int d^3\mathbf{q}\approx k_3^3\int_{0}^{2\pi}d\phi~.
\end{align}
In the squeezed limit, the most important contribution comes from $|\mathbf{p-k_1}|=k_3$ and $\mathbf{p}\approx\mathbf{k_1}$. As a result, we can calculate the integral as
\begin{align}
	I_{k_1} I_{k_2} I_{k_3}&=\frac{H^6}{k_1^3 k_2^3 k_3^3}\nonumber\\&\times 
	\left[\frac{\pi}{2}\alpha\beta\left((\frac{k_1}{k_3})^{i\mu}+(\frac{k_1}{k_3})^{-i\mu}\right)+\alpha^2A\left(\frac{k_1}{4k_3}\right)^{-i\mu}+\beta^2 A^*\left(\frac{k_1}{4k_3}\right)^{i\mu}\right]\nonumber\\&\times\left[\frac{\pi}{2}|\alpha|^2\left(\frac{k_1}{k_3}\right)^{-i\mu}+\frac{\pi}{2}|\beta|^2\left(\frac{k_1}{k_3}\right)^{i\mu}+\alpha^*\beta A^*\left(\frac{k_1}{4k_3}\right)^{i\mu}+\alpha \beta^* A\left(\frac{k_1}{4k_3}\right)^{-i\mu}\right]\nonumber\\&\times\left[\alpha^*\beta^*\pi+(\alpha^*)^2A^*\left(\frac{1}{4}\right)^{i\mu}+(\beta^*)^2A\left(\frac{1}{4}\right)^{-i\mu}\right]~.
\end{align}

Then we turn to deal with the remain angular dependent part. First of all, we simplify the product of all polarization vectors by using

\begin{align}\label{angle}
|\epsilon_{i}^{-\lambda}(\mathbf{p_1})\epsilon_{i}^{+}(\mathbf{p_2})|^2=\frac{1}{4}\left(1+\lambda\frac{\mathbf{p_1}\cdot\mathbf{p_2}}{p_1p_2}\right)^2~.
\end{align}
Under the squeezed limit, the angular part (AP) is
\begin{align}
AP=&\epsilon_{i_1}^+(\mathbf{p_1})\epsilon_{i_1}^{-\lambda}(\mathbf{k_1})\epsilon_{j_1}^+(\mathbf{k_1-p_1})\epsilon_{j_1}^{-\lambda}(\mathbf{k_1})\epsilon_{i_2}^{-\lambda}(\mathbf{-k_1})\epsilon_{i_2}^{+}(\mathbf{-p_1})\nonumber\\&\times\epsilon_{j_2}^{-\lambda}(\mathbf{-k_1})\epsilon_{j_2}^{+}(\mathbf{-k_1+p_1})\epsilon_{i_3}^{-\lambda}(\mathbf{k_3})\epsilon^{+}_{i_3}(\mathbf{p_1-k_1})\epsilon_{j_3}^{-\lambda}(\mathbf{k_3})\epsilon^{+}_{j_3}(\mathbf{k_1-p_1})~.
\end{align}
 By using equation \eqref{angle},
\begin{align}\label{AP}
AP=&\frac{1}{4}\left(1+\lambda\frac{\mathbf{k_1}\cdot\mathbf{p_1}}{k_1p_1}\right)^2\frac{1}{4}\left(1+\lambda\frac{\mathbf{k_1}\cdot(\mathbf{k_1-p_1})}{k_1(k_1-p_1)}\right)^2\nonumber\\\times&\epsilon_{i_3}^{-\lambda}(\mathbf{k_3})\epsilon^{+}_{i_3}(\mathbf{p_1-k_1})\epsilon_{j_3}^{-\lambda}(\mathbf{k_3})\epsilon^{+}_{j_3}(\mathbf{k_1-p_1})~.
\end{align}
  One way to calculate the left un-contract term is to write down the polarization vectors explicitly. Specifically, if $\mathbf{k}=(\sin\theta \cos\phi,\sin\theta
\sin\phi,\cos\theta)$, then
\begin{align}
\vec{\epsilon}_{(\pm)}=\frac{1}{\sqrt{2}}(\cos\theta \cos\phi\mp i\sin\phi,\cos\theta\sin\phi\pm i\cos\phi,-\sin\theta)~.
\end{align}
Without loss of generality, and follow the convenient choice about loop momentum configuration before, we can label momentum vectors as 

	\begin{align}
	&\mathbf{k_1}\approx-\mathbf{k_2}=(0,0,k_3)~,\\
	&\mathbf{k_3}=k_3(0,\sin \theta,\cos \theta)~,\\
	&\mathbf{k_1-p_1}=k_3\left(\frac{\sqrt{3}}{2}\cos\phi,\frac{\sqrt{3}}{2}\sin\phi,-\frac{1}{2}\right)~.	
	\end{align}
So that the polarization vectors with respect to different momentum can be labeled as
\begin{align}
&\mathbf{\epsilon^{\pm}}(\mathbf{k_3})=\frac{1}{\sqrt{2}}(\mp i,\cos \theta,-\sin \theta )~,\\
&\mathbf{\epsilon}^+(\mathbf{k_1-p})=\frac{1}{\sqrt{2}}\left(-\frac{1}{2}\cos \phi-i \sin \phi,-\frac{1}{2} \sin \phi+i \cos\phi,-\frac{\sqrt{3}}{2}\right)~.
\end{align}
Substitute these expressions into the angular part and then integrate out $\phi$. Because positive helicity is much larger than  negative helicity, here we choose $\lambda=+$ only. Then the final result of the angular part is
\begin{align}
AP_{(\lambda=+)}=\frac{\pi}{2^8}\sin ^2 \theta~.
\end{align}
Where $\theta$ here is the angle between $\mathbf{k_1}$ and $\mathbf{k_3}$. Actually, this $\sin^2 \theta$ dependence is no sensitive to our approximation. This angular dependent part will remain unchanged in a complete calculation. For example, even if we don't choose the loop momentum configuration with leaving the integration about $\theta$ part. Follow the method above, the explicit expression of the second line of ~\eqref{AP} is $\frac{\pi}{4}\sin^2\theta(3\sin^2\theta_2-2)$ where $\theta_2$ is the angle between $\mathbf{k_1-p_1}$ and $\mathbf{k_1}$. We can see that the $\sin^2\theta$ dependence still exists.

Finally, the three-point function of tensor fluctuation is
\begin{align}\label{signal}
&\langle h_{k_1}^+ h_{k_2}^+ h_{k_3}^+ \rangle=-\frac{H^6}{4Mp^6}\frac{\pi \sin^2 \theta}{k_1^6}\nonumber\\&\times\left[\frac{\pi}{2}\alpha\beta\left((\frac{k_1}{k_3})^{i\mu}+(\frac{k_1}{k_3})^{-i\mu}\right)+\alpha^2A\left(\frac{k_1}{4k_3}\right)^{-i\mu}+\beta^2 A^*\left(\frac{k_1}{4k_3}\right)^{i\mu}\right]\nonumber\\&\times\left[\frac{\pi}{2}|\alpha|^2\left(\frac{k_1}{k_3}\right)^{-i\mu}+\frac{\pi}{2}|\beta|^2\left(\frac{k_1}{k_3}\right)^{i\mu}+\alpha^*\beta A^*\left(\frac{k_1}{4k_3}\right)^{i\mu}+\alpha \beta^* A\left(\frac{k_1}{4k_3}\right)^{-i\mu}\right]\nonumber\\&\times\left[\alpha^*\beta^*\pi+(\alpha^*)^2A^*\left(\frac{1}{4}\right)^{i\mu}+(\beta^*)^2A\left(\frac{1}{4}\right)^{-i\mu}\right]  \delta^3(\mathbf{k_1+k_2+k_3}) \nonumber\\&=-\frac{H^6}{4Mp^6}\frac{\pi \sin^2 \theta}{k_1^6}\left(C_1+C_2\left(\frac{k_1}{k_3}\right)^{-2 i \mu}+C_3\left(\frac{k_1}{k_3}\right)^{2 i  \mu}\right) \delta^3(\mathbf{k_1+k_2+k_3})\\&\equiv (2\pi)^7 \delta^3(\mathbf{k_1+k_2+k_3}) \frac{B(k_1, k_3)}{k_1^6}~,
\end{align}
where we define
\begin{align}
	 B{(k_1, k_3)}\equiv\frac{-H^6 \sin^2\theta}{8 Mp^6 (2\pi)^6}\left( C_1+C_2\left(\frac{k_1}{k_3}\right)^{-2 i \mu}+C_3\left(\frac{k_1}{k_3}\right)^{2 i  \mu}\right)~.
\end{align} 
\begin{figure}[htbp]
	\centering 
	\includegraphics[width=0.6\textwidth]{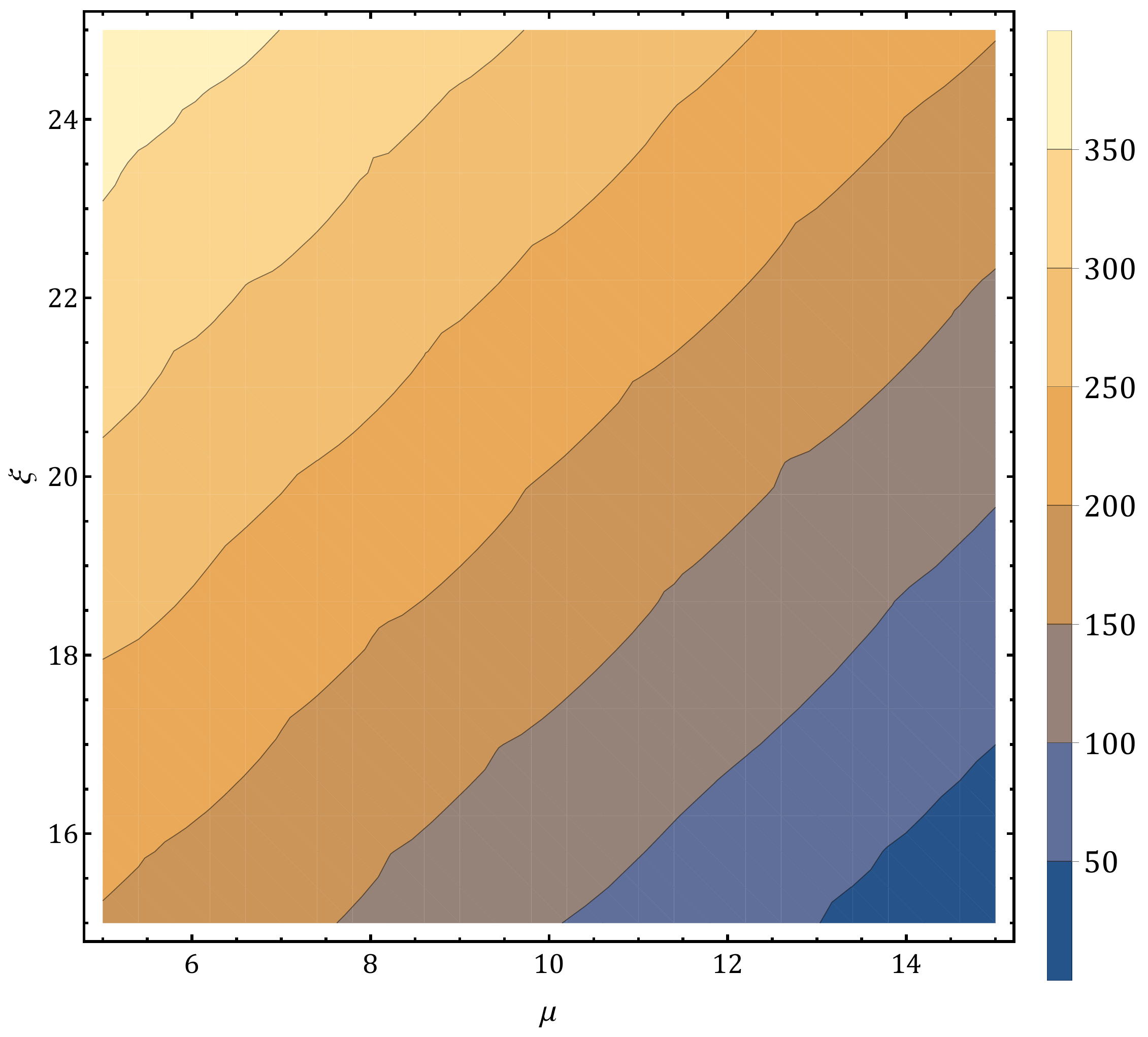}
	\caption{\label{figure1}Contour plot of Log$|C_1|$.}
\end{figure}

The different coefficients $ C_1$, $C_2$ and $C_3$ are
\begin{align}
&C_1=\left(4^{-i\mu} \left(\alpha^*\right)^2 A^*+\pi  \alpha^* \beta^*+4^{i\mu} A \left(\beta^*\right)^2\right)\nonumber\\&\times
(2^{-1+2 i\mu} \pi  \alpha^2 A \left| \beta\right| ^2+\alpha^2 A \beta \alpha^* A^*+2^{-1+2 i\mu} \pi  \alpha^2 A \beta
\beta^*+2^{-1-2 i\mu} \pi  \beta^2 \left| \alpha\right| ^2 A^*\nonumber\\&+2^{-1-2 i\mu} \pi  \alpha \beta^2 \alpha^* A^*+\alpha A \beta^2 A^*
\beta^*+\frac{1}{4} \pi ^2 \alpha \beta \left| \alpha\right| ^2+\frac{1}{4} \pi ^2 \alpha \beta \left| \beta\right|
^2)~.
\\&C_2=\left(4^{-i\mu} \left(\alpha^*\right)^2 A^*+\pi  \alpha^* \beta^*+4^{i\mu} A \left(\beta^*\right)^2\right)\nonumber\\& \left(4^{2
	i\mu} \alpha^3 A^2 \beta^*+2^{-1+2 i\mu} \pi  \alpha^2 A \left| \alpha\right| ^2+2^{-1+2 i\mu} \pi  \alpha^2 A \beta
\beta^*+\frac{1}{4} \pi ^2 \alpha \beta \left| \alpha\right| ^2\right)~.\\&
C_3=\left(4^{-i\mu} \left(\alpha^*\right)^2 A^*+\pi  \alpha^* \beta^*+4^{i\mu} A \left(\beta^*\right)^2\right)\nonumber\\& \left(4^{-2
	i\mu} \beta^3 \alpha^* \left(A^*\right)^2+2^{-1-2 i\mu} \pi  \alpha \beta^2 \alpha^* A^*+\frac{1}{4} \pi ^2 \alpha \beta \left|
\beta\right| ^2+2^{-1-2 i\mu} \pi  \beta^2 A^* \left| \beta\right| ^2\right).
\end{align}
We draw a contour plot of the coefficient $C_1$ in Figure.\ref{figure1} to see the exponential dependence on $(\xi -\mu)$ more clearly. This result is as expected, that the signal is suppressed by the Boltzmann factor $(e^{-\pi\frac{m}{H}})$ and amplified by chemical potential $(e^{\pi\xi})$.

\begin{figure}[htbp]
	\centering 
	\includegraphics[width=0.6\textwidth]{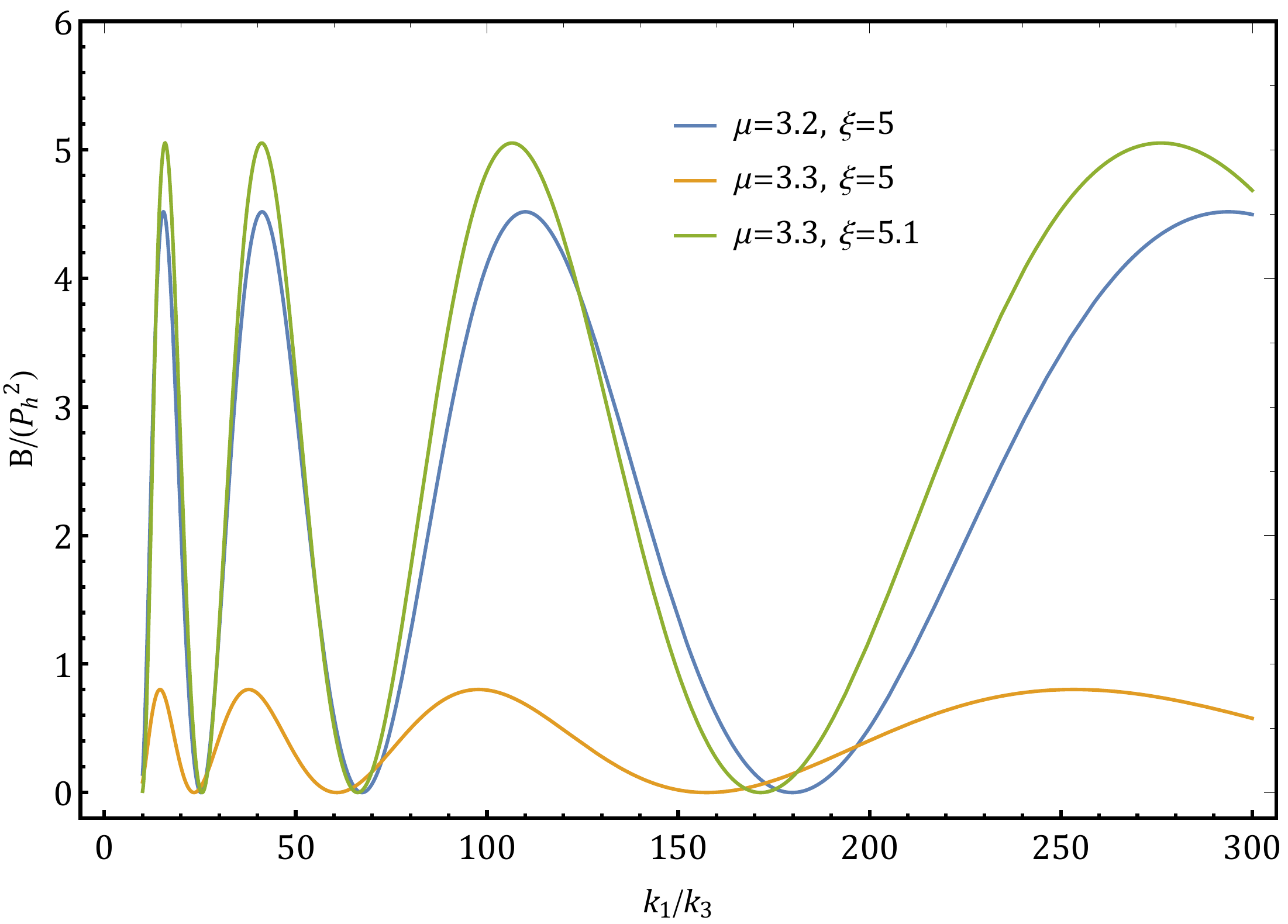}
	\caption{ \label{figure2}The oscillation bispectrum as functions of the momentum ratio~$k_1/k_3$. The blue, orange and green curves correspond to $(\mu=3.2,\xi=5), (\mu=3.3,\xi=5)$ and $ (\mu=3.3,\xi=5.1)$, respectively.}
\end{figure}

In the limit of large chemical potential, $C_1$ becomes a real number, and $C_2$ and $C_3$ are conjugate to each other. As a result, the bispectrum is real. To make more sense of the above expression,  we further expand these three coefficients by assuming $\xi \gg \mu\gg 1$. The leading contributions are
\begin{align}
	|C_1|=2|C_2|=2|C_3|\approx \mathcal{O}\left(e^{6\pi(\xi-\mu)}{\mu}^{\frac{9}{2}}\right).
\end{align}
In Figure.\ref{figure2}, we plot the bispectrum $B(k_1,k_3)/(P_h^2)$ as functions of different momentum ratio $k_1/k_3$, where $P_h$ is the normal tensor power spectrum $\frac{2 H^2}{\pi^2 {Mp}^2}$, and we choose $H/M_p=10^{-6}$ in the estimate. The blue, orange and green solid lines represent different parameter choices $(\mu=3.2,~\xi=5), (\mu=3.3,~\xi=5)$ and $ (\mu=3.3,~\xi=5.1)$. Clearly, the amplitude increases rapidly by slightly changing the chemical potential $\xi$.

However, $\xi$ cannot be too large so that the back-reaction to the inflation background is small. The constraint of the massless case is given by~\cite{Barnaby:2011qe},  as for massive vector fields case see~\cite{xitong}. In~\cite{xitong}, the mass is relatively small thus the mass term on the index is ignored. A more accurate constraint is
\begin{align}
	m H^3 e^{2 \pi (\xi-\mu)}<3 H^2M_p^2 ~.
\end{align}

In our case, $ m= \mathcal{O}(1) H$ so that the constraint is $\xi-\mu < \frac{1}{2 \pi} \ln \frac{M_p^2}{H^2}$. When ${M_p}/{H}\approx 10^6$, $\xi-\mu$ is supposed to be smaller than 4.4.
As we mentioned before, through the amplification process brought by chemical potential, GW sourced by vector fields can possibly be detected by interferometers. Here we present rough estimates of the possibility to observe three-point through interferometers such as LISA and advanced LIGO~\cite{sorbo1,Barnaby:2011qe,Domcke:2016bkh,Domcke:2017fix}. First of all, the three-point function is roughly the same order of $ \langle h h\rangle^\frac{3}{2}$ and the amplitude of the tensor perturbation is given by
\begin{align}
	\Omega_{GW}\equiv\frac{\Omega_{R,0}}{24} P\approx\frac{\Omega_{R,0}}{24} \left(\frac{2H^2}{\pi^2 Mp^2}+|B|^\frac{2}{3}\right),
\end{align}
where $\Omega_{R,0}=8.6\cdot 10^{-5}$ refers to the radiation energy density today and $P$ is the sum of power spectrum of all polarizations. Usually, one of the polarizations is much larger than the other so that we can only consider the dominate polarization. Also, to obtain the oscillation signal~\eqref{signal} does not require extremely squeezed configurations. This can be understood through the equation of motion of vector fields~\eqref{eomA}, when the mass term dominate than other terms $m^2/{(H^2 \tau^2)\gg \max(k^2,2k\xi/\tau)}$, the massive fields behavior like $(-\tau)^{i \mu}$. For generating the oscillation signal, the soft momentum should satisfy $|k_3\tau|<\min(\mu,\frac{\mu^2}{2 \xi}) $ and the hard momentum $|2 k_1 \tau| \approx \mu$. We assume $\xi > \mu$ for getting an amplified result. As a result, even if $|k_3/k_1| < \frac{u}{\xi}$, the oscillation signal can be achieved without studying extremely squeezed limits. So we choose $k_1/k_3= 5$ as a benchmark, and we also choose $H/M_p=10^{-6}$ in our estimates.
\begin{figure}[htbp]
	\centering 
	\includegraphics[width=0.7\textwidth]{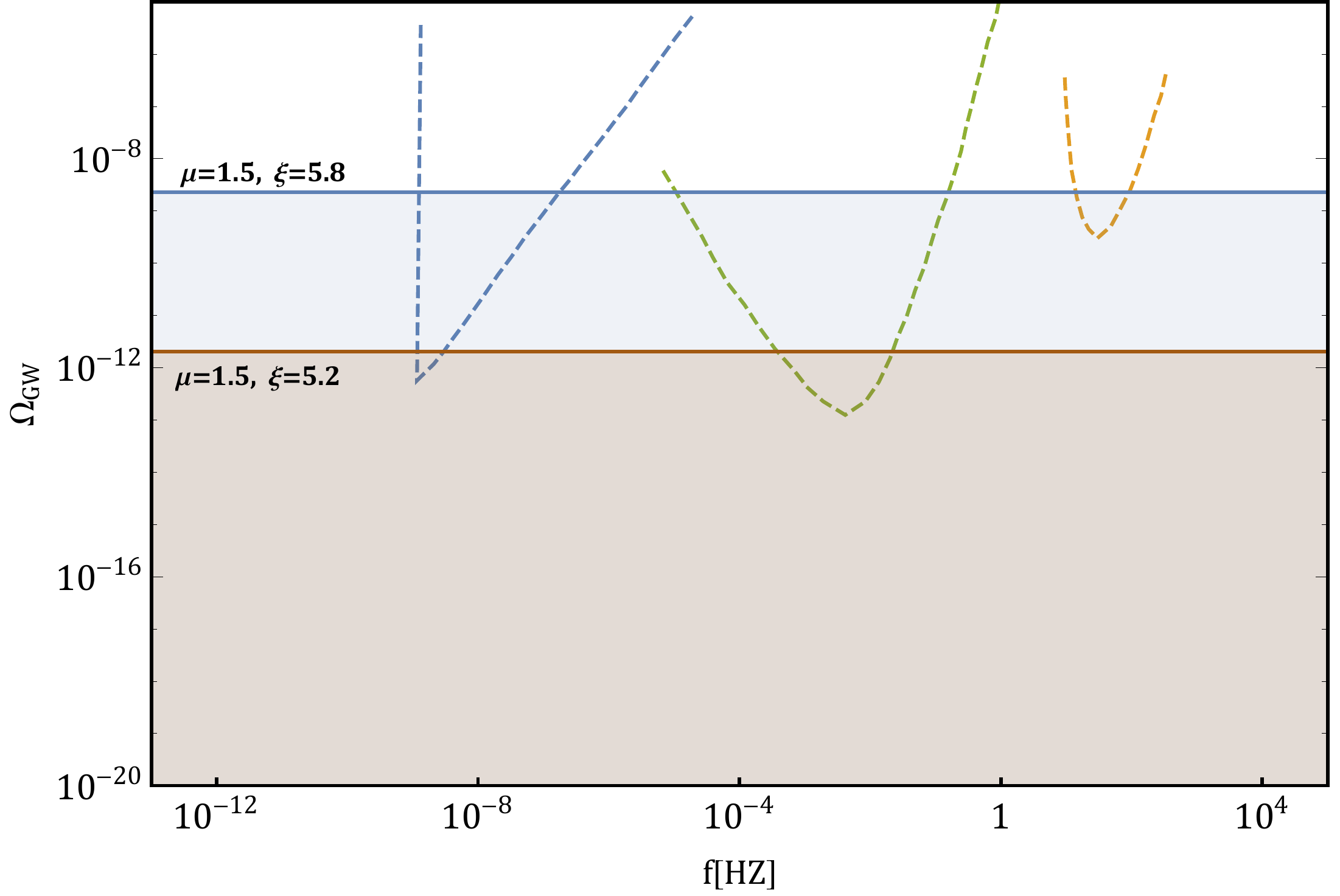}
	\caption{\label{figure3} The dashed lines represent expected sensitivities of upcoming experiments. From left to right the sensitivity curves are for the milli-second pulsar timing (blue), LISA (green) and LIGO (orange), respectively. The blue and brown solid lines correspond to amplitude estimated with parameters ($\mu=1.5,\xi=5.8$) and ($\mu=1.5, \xi=5.2$).}
\end{figure}
We plot our estimate in Figure.\ref{figure3} (follows~\cite{Domcke:2016bkh,Domcke:2017fix}) with choosing $\mu=1.5,~\xi=5.8$ and $\mu=1.5,~ \xi=5.2$ respectively. The dashed lines correspond to expected sensitivities of upcoming experiments. From left to right, they are expected sensitivity of SKA~\cite{Kramer:2004rwa}, LISA and LIGO. GW in such small scales, corresponding to late times during inflation, is not tightly constrained. However, at large scales such LSS and CMB scales, the chemical potential $\xi$ for massless vectors is required to less than 2.5 to avoid too large non-Gaussianities which was constrained by nowadays experiment. More rigorous calculations about the relationship between GW amplitude and frequency as well as the number of e-folds~($N$) requires more specific inflation models. 

\section{Conclusion}\label{conclusion}
In this work, we study the cosmological collider signal of massive vector bosons through GW. The rolling inflaton as a chemical potential can lead to particle production of massive vector fields. As a result, one polarization mode of vector fields is extremely amplified by a factor of $e^{\pi \xi}$. We calculate the three-point function of tensor fluctuation in the squeezed limit. The angular dependence and the oscillatory due to the vector mass are shown. Without a chemical potential, the signal is suppressed by Boltzmann factor $e^{-\pi \mu}$ and thus hard to observe in future experiments. Thanks to the particle production process, this signal is amplified by a chemical potential related factor $e^{\pi \xi}$. When $\xi$ is larger than $\mu$, the signal can be much larger and may be detected through the future upcoming interferometer experiments. At small scales, corresponding to late times during inflation there are fewer observational constraints. We estimate the amplitude of this signal and show that it is possible to detect through GW interferometer experiments.  

There are a number of interesting questions for future studies. In the calculation, we have used some approximations. It is interesting to find a more precise and efficient way for both analytic and numerical computations. In addition, more detailed studies require specific inflation models and to study the evolution of chemical potential $\xi$ with e-folding number. Also $\xi$ at CMB scales needs careful consideration to avoid generating too large features on the power spectrum or non-Gaussianities violating nowadays experiments constraints. To observe the signal at interferometer scales, signal to noise also need to be careful considered~\cite{Bartolo:2018qqn}. It is also interesting to generalize the study to higher spin massive particles, as possible indications of string theory.

\acknowledgments
We thank Qianhang Ding, Xi Tong and Zhong-Zhi Xianyu for useful discussions. This work is supported in part by GRF Grants 16301917, 16304418 and 16303819 from the Research Grants Council of Hong Kong.

\end{document}